\documentclass[twocolumn]{emulateapj}
\usepackage{epstopdf}
\usepackage{graphicx} 
\usepackage{color} 

\renewcommand{\b}[1]{\mbox{\boldmath{$#1$}}} 

\newcommand{\be}{\begin{eqnarray}}
\newcommand{\ee}{\end{eqnarray}}
\newcommand{\eq}[1]{eq.~(\ref{eq:#1})}

\newcommand{\grad}{\mbox{\boldmath{$\nabla$}}}

\newcommand{\shat}{\b{\hat{s}}}

\begin{document}

\title{Photon Feedback: Screening and the Eddington Limit}
\author{Aristotle Socrates$^{1,}$\altaffilmark{3} and Lorenzo Sironi$^{2,}$\altaffilmark{4}}
\altaffiltext{3}{John N. Bahcall Fellow: socrates@ias.edu}
\altaffiltext{4}{NASA Einstein  Fellow: lsironi@cfa.harvard.edu }

\affil{$^{1}$Institute for Advanced Study, Einstein Drive, Princeton, NJ 08540\\
$^{2}$Harvard-Smithsonian Center for Astrophysics, 60 Garden St., Cambridge, MA 02138}
\begin{abstract}
Bright star-forming galaxies radiate well below their Eddington Limit.  The value of the
flux-mean opacity that mediates the radiation force onto matter is orders of magnitude
smaller than the UV or optical dust opacity.  On empirical grounds, it is shown that  
high-redshift ULIRGs radiate at two orders of magnitude below their Eddington Limit, while the local starbursters
M82 and Arp~220 radiate at a few percent of their Eddington Limit.  A simple model for the radiative transfer of UV and optical light in dust-rich environments is considered.  Radiation pressure
on dust does not greatly affect the large-scale gas dynamics of star-forming galaxies.  
\end{abstract}

\section{Introduction}

Star formation is inefficient.  The timescale at which gas is converted
into stars is long in comparison to the gravitational collapse time.  In order to 
explain this, some form of self-regulation or ``feedback" is often invoked.  

There are many different mechanisms:  core-collapse supernovae, thermonuclear supernovae, radiation 
pressure on dust, cosmic ray pressure, stellar winds, heating due to ionizing radiation, jets
from proto-stellar disks, turbulent stresses, magnetic stresses, ram-pressure
stripping, pulsars winds, cosmic ray heating, dark matter 
annihilation as well as the various forms of 
energy release from super-massive black hole growth (see the Appendix for a brief literature review).

The UV and optical radiation force that directly results from star formation seems to be 
a promising candidate and has recently gained a significant amount of attention.  Second
only to prompt neutrinos from core-collapse supernovae, starlight is the dominant 
source of energy release in the Universe.  The nuclear binding energy that efficiently fuels
starlight is orders of magnitude larger than the gravitational binding energy of even the 
most massive galaxies.  In addition, UV and optical starlight is extremely well coupled
to the interstellar medium via scattering and absorption onto dust grains.  An assessment of this 
mechanism is the subject of what follows.

\section{the Eddington Limit}\label{s: Eddington}

The radiation force density is given  by (cf. Blaes \& Socrates 2003; Rybicki \& Lightman 1979)
\be
\frac{\rho}{c}\int d\nu \, \kappa_{\nu} \,{\bf F}_{\nu}~,
\ee
where $\kappa_{\nu}$ is the the sum of scattering and absorption opacities at photon 
frequency $\nu$, ${\bf F}_{\nu}$ is the radiative flux and $\rho$ is the fluid mass density.
By equating this to the  gravitational force and integrating over the galactic surface we 
arrive at the Eddington Limit
\be
L_{_{\rm Edd}}=\frac{4\pi\,G\,M_{\rm enc}\,c }{\kappa_{_F}}~,
\ee
where $M_{\rm enc}$ is the enclosed dynamical mass and  
$\kappa_{_{F}}$ is the flux-mean opacity, which can be defined as 
\be
\kappa_{_F}\equiv \frac{\int d\nu\,\kappa_{\nu}\,L_{\nu}}{\int d\nu\, L_{\nu}}~,
\ee
if the Spectral Energy Distribution (SED) and the opacity law are uniform on the 
surface of the galaxy, so that
\be
\int d{\bf A}\cdot\int d\nu\, \kappa_{\nu}\,{\bf F}_{\nu}=\int d\nu\int d{\bf A}\cdot
\kappa_{\nu}\,{\bf F}_{\nu}=\kappa_{_F}\,L~.
\ee
Note that the Eddington Limit is a statement of hydrostatic balance that is
independent of geometry and whether or not the flow is optically thick or thin 
 (Socrates 2012; Abramowicz et al. 1980).

The most luminous starburst galaxies in the Universe have luminosities 
of order $L_{\rm max}\lesssim 10^{47} $ ergs/s.  The starlight couples to gas by scattering 
and absorption onto dust grains.  The UV opacity on dust is $\kappa_{_{\rm UV}}
\sim 10^3\kappa_{es}$, where $\kappa_{es}$ is the electron scattering opacity.  
The most massive galaxies in the Universe possess an enclosed mass of order
$M_{\rm enc}\simeq 10^{12}M_{\odot}$.  With this combination of $L_{{\rm max}}$,  $\kappa_{_{\rm UV}}$
and $M_{\rm enc}$, it seems as though $L_{\rm max}\approx L_{_{\rm Edd}}$, when choosing $\kappa_{_F}\sim \kappa_{_{\rm UV}}$.  This serves
as the central argument behind the Murray et al. (2005) photon feedback model.  

Such an argument is appealing.  It provides a simple and compelling theoretical framework for 
understanding the coupling of the various forms of energy release that accompany 
star formation with the gravitational content of the galaxy in question.  Ultimately, many of 
the principles outlined in Murray et al. (2005) may contribute to our understanding 
of the Faber-Jackson (1976) relation, the fundamental plane of elliptical galaxies, and 
the maximum luminosity of star-forming galaxies.
 
\begin{figure}[t]
\epsscale{1.25}
\plotone{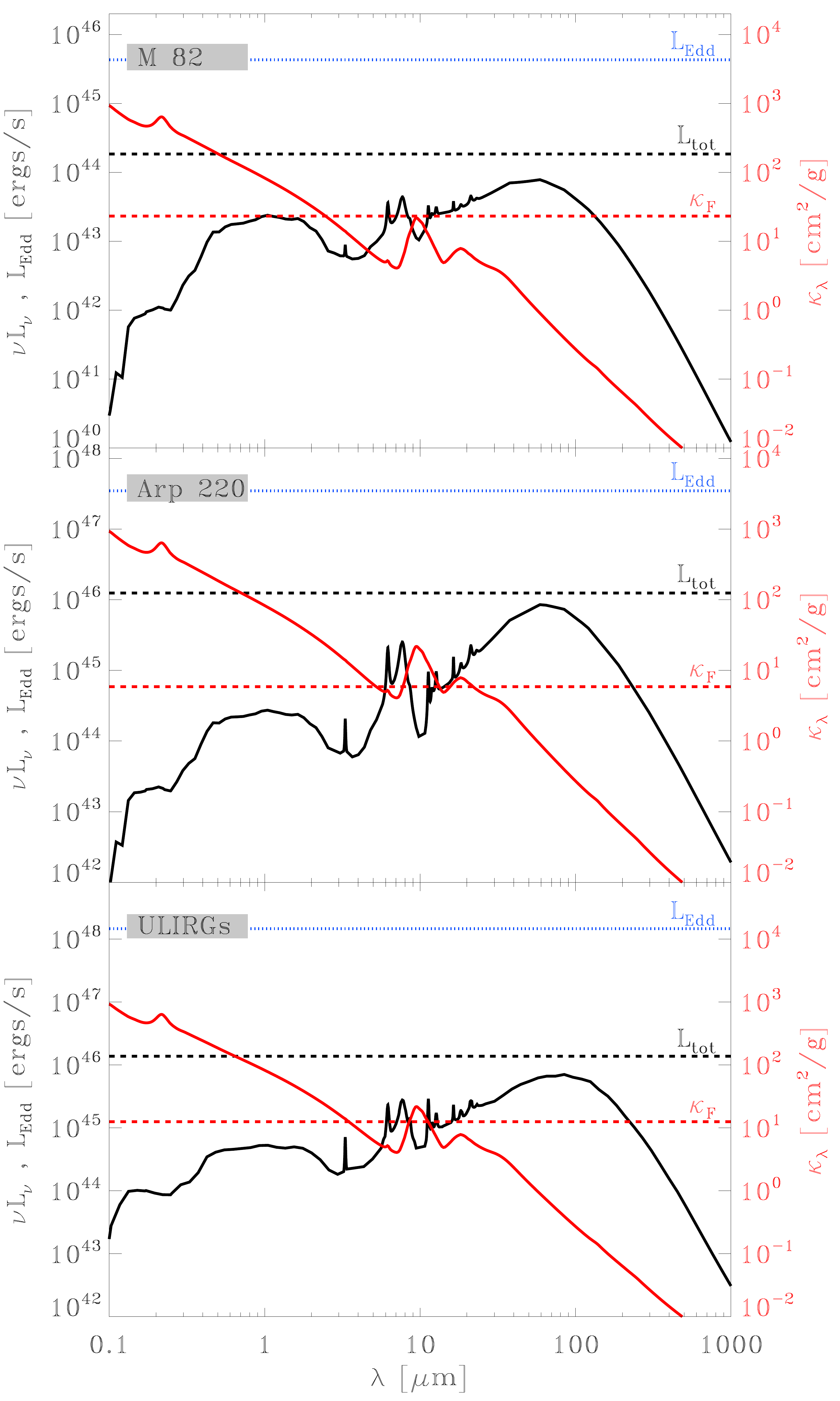}
\caption{SED (black solid line) and total luminosity (black dashed line), wavelength-dependent opacity $\kappa_\lambda$ (red solid line, including both scattering and absorption) and flux-mean opacity (red dashed line) and 
the Eddington Limit (blue dotted line) for luminous star-forming galaxies. For M82 and Arp 220, the SED is taken from Silva et al. (1998). The enclosed dynamical mass is estimated as $M_{\rm enc}\sim2\times 10^{9}M_\odot$ for M82 (Greco et al.  2012) and $M_{\rm enc}\sim4\times 10^{10}M_\odot$ for Arp 220 (Silva et al. 1998). The average SED and median enclosed mass for ULIRGs are taken from the sample of Micha{\l}owski et al. (2010). The opacity law is 
assumed to follow that of the Milky Way as calculated by Weingartner \& Draine (2001) for $R_V=3.1$ and a gas-to-dust mass ratio $f_{gd}=100$. The opacity model by Semenov et al. (2003) yields similar results.}
\label{fig: sed}
\end{figure}

However, $\kappa_{_{\rm UV}}\neq \kappa_{_F}$.  In fact, for starbursting galaxies
$\kappa_{_F}\ll \kappa_{_{\rm UV}}$.   In Figure \ref{fig: sed}, the frequency, or wavelength, dependent dust opacity (solid red line) is super-imposed upon 
the SED of the compact -- presumably -- starbursting galaxies Arp~220 and M82, and upon an average spectrum of high-$z$ ULIRGs (black solid line).  The flux-mean 
opacity $\kappa_{_F}\sim 5-20\,{\rm cm^2/g}$ for all of the sources (red dashed line).  Bright star-forming galaxies 
radiate well below their photon Eddington Limit (blue dotted line). 

The primary cause of this shortfall in the radiation force is that  most of the light
belongs to wavelengths where the opacity is small.  Or conversely, there is relatively little
light at frequencies where the opacity is large.  

The situation is similar to the radiative
transfer in a stellar envelope where the opacity is large for values of photon energy 
near atomic transitions.  Yet, photons in stellar envelopes 
escape primarily by  avoiding such  regions of high opacity, which leads to deep 
absorption features.  Consequently, the flux-mean opacity in  stellar envelopes 
remains close to the low continuum value.

\begin{figure}[t]
\epsscale{1.2}
\plotone{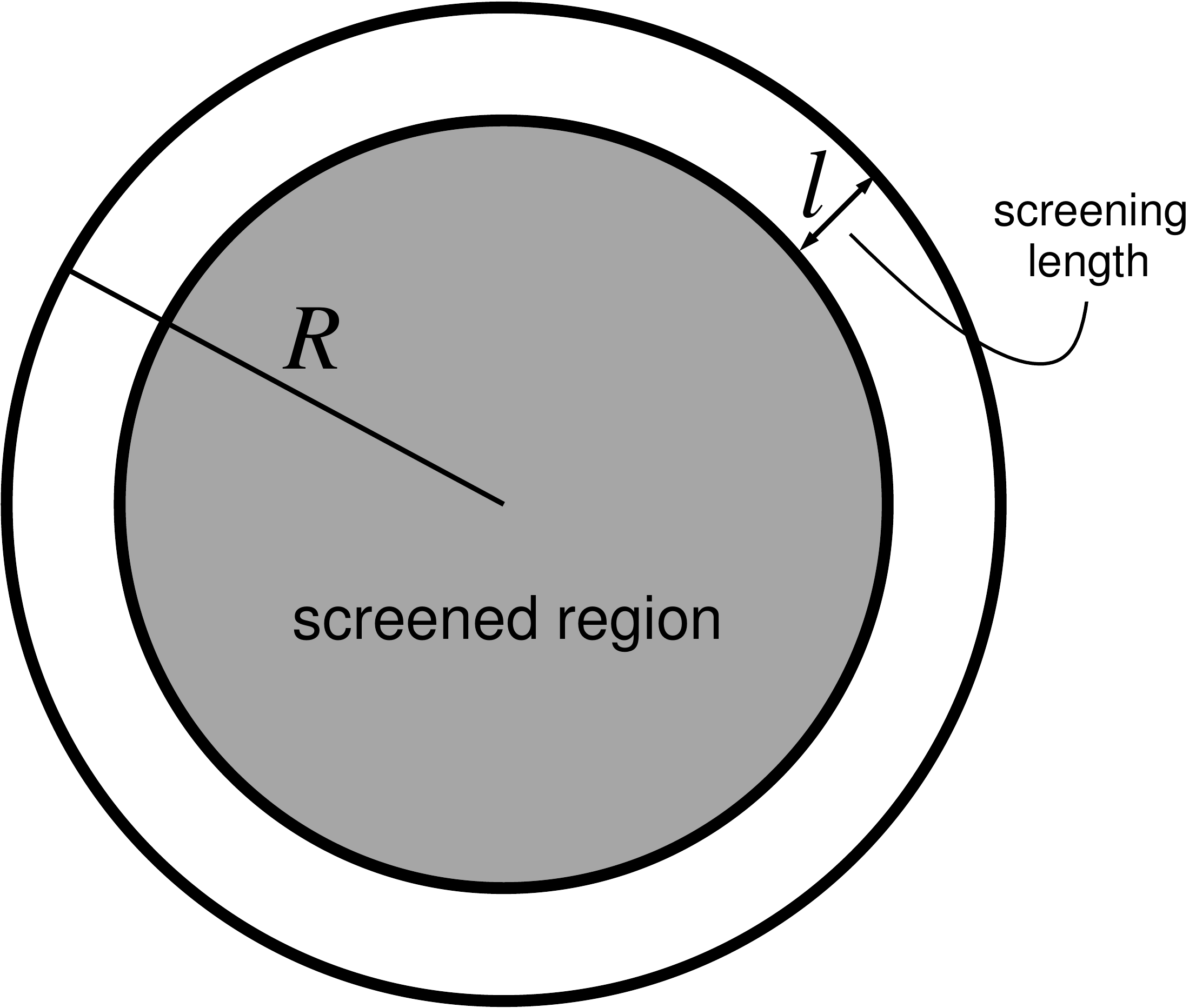}
\caption{A simple model of a star-forming galaxy.  The starlight is assumed to be injected uniformly
within the radius $R$.  Due to the large optical depth at UV/optical wavelengths, the cold gas in the bulk of the galaxy 
 screens the starlight except for an outer region whose thickness is of order 
the screening length $l$.  The bolometric output generated from within the galaxy ultimately 
escapes in the form of quasi-thermal FIR radiation.  Similar arguments for explaining the 
ratio of UV/FIR power can be made in cylindrical geometry as well, with the same 
result. }
\label{fig: screen}
\end{figure}

\section{UV/optical screening}\label{s: screen}

Figure \ref{fig: sed} indicates that the ratio of the UV to FIR luminosity in bright star-forming galaxies is low -- of order a few percent.  As a result, 
the flux-mean opacity $\kappa_{_{F}}$ is correspondingly diminished and the galaxy 
in question radiates well below its Eddington value.  

An increase in the gas surface density leads to a corresponding increase in the rate of star formation.
Therefore, dense gas serves both as a source of UV/optical starlight radiation pressure via star formation, while simultaneously screening itself via absorption and re-processing of the UV light ultimately into
the FIR.  A depiction of the UV/optical screening in a dense star-forming galaxy is shown in Figure \ref{fig: screen}.   

Consider a galaxy of characteristic length scale $R$, with uniform density, temperature $(\rho ,T)$
and composition.  Under these assumptions, the 
starlight screening length $l$ is a constant.  For a constant value of the 
star formation rate per unit volume, the ratio of UV light to FIR radiation is 
\be
\frac{L_{_{\rm UV}}}{L_{_{\rm FIR}}}\sim\frac{l_{_{\rm UV}}}{R}~.
\label{e: screen}
\ee
The expression above may be a good approximation for cool dust-rich galaxies that 
are opaque to starlight, where the FIR luminosity 
$L_{_{\rm FIR}}$ is responsible for most of the bolometric energy release.  

The simple picture outlined above may explain the small ratio of UV to FIR photon 
power in starbursting galaxies.  An attempt to quantify the properties of starlight screening by
considering the equation of radiative transfer is given below.

\subsection{Radiative Transfer}

The spatial and directional dependence of the starlight is  
quantified by the specific intensity $I_{\nu}$.  Dust grains are the 
primary source of scattering and absorption of starlight, where absorption acts
to redistribute the input stellar photon energy longwards  to IR and FIR frequencies.  At these
wavelengths, the dust opacity is relatively small and the radiant energy can more easily 
escape.

In steady state, the equation of radiative transfer for the specific intensity $I_\nu(\b x,\b{\hat{s}})$ along a direction $\shat$ reads (Rybicki and Lightman 1979)
\be\label{eq:diff0}
\grad\cdot [I_\nu(\shat) \shat]=-(\alpha_\nu+\sigma_\nu)I_\nu(\shat)+ \nonumber\\
+\sigma_\nu\! \int_{4 \pi} I_\nu(\b{\hat{s}}') f_\nu(\shat\cdot \b{\hat{s}}') d\Omega'+j_\nu(\shat)
\ee
where the spatial dependence on $\b x$ has been suppressed. Here,  $\alpha_\nu(\b x)$ is the absorption coefficient and $\sigma_\nu(\b x)$ the scattering coefficient. They are, respectively, 
equal to the inverse of the mean free paths for absorption and scattering of photons
of frequency $\nu$. The emission coefficient, or source,  is given by $j_\nu(\b x, \shat)$
and the normalized differential scattering cross section, $f_\nu(\shat\cdot \b{\hat{s}}')$, satisfies
\be
\!\int_{4 \pi}f_\nu(\shat\cdot \b{\hat{s}}') d\Omega'=1~.
\ee
By defining the mean intensity $J_\nu(\b x)=\!\int I_\nu (\b x,\b{\hat{s}})\,d\Omega$,  the radiative flux ${\bf F}_\nu(\b x)=\!\int I_\nu (\b x,\b{\hat{s}}) \,\shat\, d\Omega$ and the mean emissivity $q_\nu(\b x)=\!\!\int j_\nu (\b x,\b{\hat{s}})\,d\Omega$,\footnote{In the following, the source, quantified by the emission coefficient $j_\nu$, is taken 
to emit isotropically.} the  
radiative transfer equation averaged over the propagation angle reduces to 
\be\label{eq:diff1}
\grad \cdot {\bf F}_\nu=-\alpha_\nu J_\nu +q_\nu~.
\ee
In the diffusion approximation, the specific intensity $I_\nu$ can be expressed as an isotropic part (i.e., the mean intensity $J_\nu$) plus a small directional flux ${\bf F}_\nu$.  That is, 
\be
I_\nu(\shat)=\frac{1}{4\pi} J_\nu+\frac{3}{4\pi} {\bf F}_\nu\cdot \shat
\ee
where the factor of three on the right hand side results from enforcing the Eddington closure 
approximation.
Substitution of this expression into \eq{diff0},  multiplying by $\shat$ and integrating over all solid angles, leads to a relation between the mean intensity and the flux of 
the following form
\be\label{eq:diff2}
{\bf F}_\nu=-D_\nu \grad J_\nu
\ee
where
\be\label{eq:DD}
D_{\nu}(\b x)\equiv \frac{1}{3 [(1-g)\sigma_\nu+\alpha_\nu]}=\frac{1}{3\,\rho\, \kappa_\nu}
\ee
is the photon diffusion coefficient, and $g$ is the average cosine of the scattering angle. The two relations in eqs.~(\ref{eq:diff1}) and (\ref{eq:diff2}) between the mean intensity and the flux can be combined to give a diffusion equation for the mean intensity alone
\be
\grad \cdot (D_\nu \grad J_\nu)-\alpha_\nu J_\nu= -q_\nu~.
\ee

In a uniform medium,  the diffusion coefficient $D_\nu$ is independent of 
position, and under these conditions
\be\label{eq:diff4}
D_\nu\grad^2 J_\nu-\alpha_\nu J_\nu= -q_\nu~,
\label{e: diff_eq}
\ee
which is in the form of a screened Poisson's equation with screening length $l_\nu\equiv\sqrt{D_\nu/\alpha_\nu}$.  The Green's function $G_{\nu}(\b x,\b x')$ for the screened Poisson's equation 
satisfies $\grad^2 G_\nu-G_\nu/l_\nu^2=-\delta(\b x-\b x')$  and is given by
\be
G_\nu(\b x, \b x')= \frac{\exp(-|\b x-\b x'|/l_\nu)}{4 \pi |\b x-\b x'|}
\ee
which is also referred to as the Yukawa potential.  Under the assumptions
that the radiation field is quasi-isotropic and photon diffusion proceeds through a uniform medium, the analogy 
of the radiative transfer problem with electrostatics is clear.  The average intensity $J_{\nu}$, 
which is equivalent to the photon energy density, can be thought of as an electrostatic
potential that is produced by some source of radiation $q_{\nu}$, which itself can be 
thought of as an electrostatic charge density.  Furthermore, it follows that the 
radiative flux ${\bf F}_{\nu}$, responsible for the radiation force, is analogous to 
an electric field.  Absorption attenuates the mean intensity in starlight, which is therefore 
evanescent in the bulk of the galaxy,
and its effect may be thought of as a screening. 

The solution to the radiative transfer of starlight under these circumstances 
further motivates the arguments behind Figure \ref{fig: screen} and eq.~(\ref{e: screen}).

\subsection{Boundary Value Problem in Spherical Symmetry}

In a uniform sphere, the source function $q_{\nu}=q_0$ is a constant.
The inhomogeneous solution for $J_{\nu}$ to the screened diffusion equation must be regular 
at the center of the galaxy and satisfy an appropriate boundary condition at the surface $r=R$. 
For the latter, we employ the two-stream approximation (Rybicki and Lightman 1979), which relates the flux ${\bf F}_{\nu}$
to the mean intensity $J_{\nu}$ in such a way as to maintain the Eddington approximation for
the starlight.
With this, the mean intensity inside the spherical galaxy is given by
\be\label{eq:in1}
J_\nu=\frac{q_0 \,l_\nu^2}{D_\nu}\left[1-\frac{R}{r}\frac{\sinh(r/l_\nu)}{C}\right]~,
\ee
where the constant $C=C(R)$ is determined by the boundary condition at 
$r=R$ and is written as
\be
\!\!\!C\!\equiv\! \sqrt{3D_\nu \alpha_\nu} \cosh\!\left(\frac{R}{l_\nu}\right)\!\!+\!\!\left(1\!-\!\sqrt{3} \frac{D_\nu}{R}\right) \sinh\!\left(\frac{R}{l_\nu}\right).
\ee
For $l_\nu\ll R$, the mean intensity inside the sphere is almost constant and equal to $J_\nu\sim q_0 l_\nu^2/D_\nu\sim q_0/\alpha_\nu$, as expected for a homogeneous source with emissivity $q_0$ that is absorbed locally with absorption coefficient $\alpha_\nu$.

The radial component of the radiative flux $F_{\nu}=\b{\hat r}\cdot \bf F_{\nu}$ is given by 
\be
F_{\nu}=\frac{q_0 \,R \,l_\nu}{r^2}\left[\frac{r\cosh(r/l_\nu)-l_\nu \sinh(r/l_\nu)}{C¨}\right]~,
\ee 
which vanishes in the limit $r\rightarrow 0$, as expected.  Near $r=0$, $F_{\nu}$ rises linearly 
with $r$ up to $r\approx l_{\nu}$, at which point $F_{\nu}$ increases, approximately, exponentially
with scale length $l_{\nu}$ up until $r=R$, where it reaches its maximum value.
Since the UV radiation force is proportional to ${\bf F}_{\nu}$, it follows that, in the
limit $l_{_{\rm UV}}/R\approx L_{_{\rm UV}}/L_{_{\rm FIR}}\ll 1$, {\it throughout the 
bulk of the flow the UV radiation force, effectively, vanishes.}  Consequently, the so-called
``single-scattering"  approximation for UV photons (Murray, Quataert \& Thompson 2005; 2010; cf. Hopkins
et al. 2012 a,b,c) significantly -- or exponentially -- over-estimates the UV radiation force throughout the  
bulk of a given star forming flow.

In the absence of absorption (set $\alpha_{\nu} =0$ in \eq{diff4}), while still enforcing
the two-stream approximation at the surface, the un-screened intensity $J^{(0)}_{\nu}$
is given by
\be
J^{(0)}_{\nu}=\frac{q_0}{6 D^{(0)}_{\nu}}(R^2-r^2)+\frac{q_0 R}{\sqrt{3}}
\ee
where $D^{(0)}_{\nu}=[3(1-g)\sigma_\nu]^{-1}\sim 2 D_\nu $ for isotropic diffusion (i.e., $g=0$) and assuming
equal mean free paths for scattering and absorption (i.e., $\sigma_\nu=\alpha_\nu$). The corresponding 
un-screened radiative flux is then $F^{(0)}_{\nu}=q_0 r/3$.

For $R\gg \alpha_\nu^{-1}$ and $R\gg D_\nu$ (which also implies $R\gg l_\nu$), the ratio between 
screened and un-screened mean intensities at the center (i.e., at $r=0$) is
\be
 \left.\frac{J_{\nu}}{J^{(0)}_{\nu}}\right|_{r=0}=\frac{6 D^{(0)}_{\nu}}{D_{\nu}}\frac{l_\nu^2}{R^2}\sim \frac{12\, l_\nu^2}{R^2}.
\ee
It follows that, {\it for galaxies such as Arp~220 and ULIRGs whose starlight is highly screened, the 
UV/optical radiation pressure is strongly suppressed at depth.}

The ratio of outward radial fluxes at the outer boundary of the galaxy (i.e., at $r=R$)
reads
\be
\left.\frac{F_{\nu}}{F^{(0)}_{\nu}}\right|_{r=R}
 =\frac{3 l_\nu}{R}\frac{1}{\sqrt{3D_\nu\alpha_\nu}+1}\sim \frac{l_{\nu}}{R}~,
\ee
which  corroborates the intuition that led to Figure \ref{fig: screen}.
 
%


\section{Discussion and Summary}

In \S\ref{s: Eddington} an empirical method -- the calculation of the Eddington Limit -- for determining 
the relative strength of the radiation force on the surface of galaxies is described.  It is demonstrated 
that light resulting from 
star formation is too dim to affect the large-scale gas dynamics of actively 
star-forming galaxies.  

The dust opacity 
is large at small wavelengths, where the emergent spectrum is relatively dark, while 
the dust opacity is small at the long photon wavelengths, where the emergent spectrum 
is bright.  By combining these two properties it is apparent that the radiation force is, 
in general, unexpectedly weak in gas-rich star-forming sources.

Galaxies, such as the Milky Way, that are not as rich in cool dense molecular gas 
do not suffer from the same level of self-shielding that renders radiation pressure ineffective
in  starbursting sources.  However, the lack of molecular fuel also implies a relatively low level 
of star formation.  Although the flux-mean opacity may be higher, due to a harder spectrum, 
the total luminosity is low.  So, it is unlikely that un-screened galaxies radiate above
their Eddington Limit. 

The radiative transfer of starlight in luminous star-forming galaxies can be 
modeled in the diffusion approximation, including the effect of  absorption by  dust grains.  
Dust-screening greatly diminishes the value of the starlight intensity, or pressure, 
at depth.  Only an outer shell with thickness equal to the screening length contributes to the outgoing
flux of UV/optical starlight.  An increase in the density of molecular gas increases the rate of 
starlight production and the Eddington Limit may be approached.  Yet,
increasing the fuel supply necessarily leads to an increase in dust-screening, with 
corresponding downward departures from the Eddington Limit, due to the 
corresponding decrease in the flux-mean opacity.

Since the sources under consideration radiate well below their Eddington Limit, the large-scale radiation force due to FIR photons is also small in comparison to the galaxy-scale gravitational
force at the surface.  It is therefore unlikely that the FIR radiation pressure will be large
enough to be dynamically important at depth.  In the diffusion approximation, radiation pressure
is proportional to optical depth, which is proportional to column density.  Since gas pressure
is also proportional to column density, the ratio between gas and FIR radiation pressure 
should remain approximately a constant with depth.

There may be some class of galaxies in the Universe or objects within luminous star-forming galaxies
themselves (e.g., giant molecular clouds), where starlight pressure on interstellar dust leads 
to a radiation force that exceeds the local source of gravity (Murray, Quataert \& Thompson 2010).  It may 
be useful to empirically determine the Eddington Limit for these objects with the method 
described in \S\ref{s: Eddington}.

\acknowledgements
A.S. is supported by a 
John N. Bahcall Fellowship Fellowship at the Institute for Advanced Study, Princeton. L.S. is supported by NASA through Einstein
Postdoctoral Fellowship grant number PF1-120090 awarded by the Chandra
X-ray Center, which is operated by the Smithsonian Astrophysical
Observatory for NASA under contract NAS8-03060.  We gratefully thank M.~Micha{\l}owski for providing the SED templates of high-redshift ULIRGs,
as well as Jenny Greene and Avi Loeb for comments.


\appendix
\section{Review of Radiation Feedback}

In recent years, the idea that radiation pressure due to starlight 
interacting with dust grains may be comparable to gravity was considered 
by Scoville (2003).  Martin (2004) and Murray et al. (2005; MQT) hypothesized
that radiation pressure on dust grains may be responsible for driving large-scale
galactic winds in analogy to line-driven stellar winds of massive stars. MQT utilize
the large UV/optical opacity on dust to justify that, despite the low light to mass ratio 
of the galaxies, it is possible for them to radiate with super-Eddington luminosities.  
Thompson et al. (2005) focus on the effects of coupling the IR and FIR radiation to 
dust and thus, to the gaseous component, in an attempt to construct a 
starlight-powered galactic disk solution.  More recently, Murray et al. (2010)
changed focus from large scales to the scale of giant molecular clouds (GMCs), where
they posit that fully-populated star clusters born in GMCs unbind the site of their birth
once the star to gas ratio becomes large.  In a series of recent papers (Hopkins 
et al. 2011; Hopkins et al. 2012a,b,c; Faucher-Giguere et al. 2013), both the GMC-scale
starlight-driven hypothesis as well as an approximate formulation of the large-scale 
radiation force are utilized in a number of massive numerical simulations meant 
to describe a variety of star-forming galaxies.  They find that the radiation force 
is capable of significantly altering the gas dynamics of star-forming galaxies. 

Recently, Andrews \& Thompson (2011) and Zhang \& Thompson (2012) note
that, at face value, star-forming galaxies are sub-Eddington with respect to the 
Rosseland mean opacity for dust.  By taking the Rosseland mean, they assume
a functional form for the SED of the galaxy.  As a result of this, they are led to 
believe that there is some level of theoretical ambiguity when assessing the strength of the
radiative forcing, particularly when discerning between optically thick and optically thin 
regimes.  One of the most attractive features of the Eddington limit is that 
it is an empirical tool.  In \S\ref{s: Eddington} the observed SED of galaxies is integrated
against a typical dust opacity law in order to obtain the strength of the radiative 
forcing on the surface of a given galaxy, with relatively little ambiguity.  Furthermore, 
the Eddington Limit is independent of whether or not the flow is optically thick or thin.

In describing the feedback of cosmic ray protons during star formation, Socrates
et al. (2008; see also Sironi \& Socrates 2010 for the case of quasars) attempt to 
draw a distinction between cosmic ray feedback and the feedback from starlight.  In their
appendix, they note that UV photons decouple and are reprocessed into 
the FIR close to the massive star of their origin.  Once starlight is converted
into the FIR band, they argue that it is unlikely to affect the dynamics of the interstellar medium.
A critical re-examination of photon feedback is the subject of many recent numerical 
studies (Novak et al. 2012; Krumholz \& Thomson 2012;  Wise et al. 2012; 
Agertz et al. 2012; Jiang et al. 2013).  They all concur, to varying degrees 
and for various reasons, with Socrates et al. (2008) in that
that starlight pressure may not be efficient in its coupling to the interstellar
medium.  However, to solve the full radiation hydrodynamical problem of photon 
feedback is incredibly challenging from a numerical perspective due to large
ratio of the photon diffusion to gas sound speed (cf. Fernandez \& Socrates 2012).  In 
\S\ref{s: Eddington} an attempt is made to circumvent this debate by empirically determining
the strength of the radiative forcing on the surface of star-forming galaxies.

\end{document}